\documentclass[aps,prl,twocolumn,showkeys,superscriptaddress]{revtex4}
\usepackage{amssymb}
\usepackage{graphicx}
\usepackage{xcolor}
\usepackage{amsmath}
\usepackage{amsfonts}
\usepackage{amssymb}
\usepackage{braket}

\begin{document}

\title{Relaxation of the Excited Rydberg States of Surface Electrons on Liquid Helium}

\author{Erika Kawakami}
\email[E-mail: ]{e2006k@gmail.com}
\affiliation{Quantum Dynamics Unit, Okinawa Institute of Science and Technology, Tancha 1919-1, Okinawa 904-0495, Japan}
\affiliation{PRESTO, Japan Science and Technology (JST), Kawaguchi, Saitama 332-0012, Japan}
\author{Asem Elarabi}
\affiliation{Quantum Dynamics Unit, Okinawa Institute of Science and Technology, Tancha 1919-1, Okinawa 904-0495, Japan}
\author{Denis Konstantinov}
\email[E-mail: ]{denis@oist.jp}
\affiliation{Quantum Dynamics Unit, Okinawa Institute of Science and Technology, Tancha 1919-1, Okinawa 904-0495, Japan}
\date{\today}

\begin{abstract}
We report the first direct observation of the decay of the excited-state population in electrons trapped on the surface of liquid helium. The relaxation dynamics, which are governed by inelastic scattering processes in the system, are probed by the real-time response of the electrons to a pulsed microwave excitation. Comparison with theoretical calculations allows us to establish the dominant mechanisms of inelastic scattering for different temperatures. The longest measured relaxation time is around 1~$\mu$s at the lowest temperature of 135 mK, which is determined by the inelastic scattering due to the spontaneous two-ripplon emission process. Furthermore, the image-charge response shortly after applying microwave radiation reveals interesting population dynamics due to the multisubband structure of the system.    
\end{abstract}

\maketitle

Electrons trapped on the surface of liquid helium provide us with a unique two-dimensional electron system (2DES)~\cite{Monarkha-book,Andrei-book}. The bound states of the electron motion perpendicular to the surface are formed due to, on one hand, the attraction of the  electron to a weak image-charge inside the liquid and, on the other hand, the hard-core repulsion from the helium atoms, which prevents the electron from entering the liquid. The Rydberg states of such confined one-dimensional (1D) motion share certain similarities with the atomic spectrum of hydrogen ~\cite{GrimPRL1974,GrimPRB1976,LambPRL1980,VolJETP1981,CollPRL2002}. At the same time, these surface electrons (SE) can move freely parallel to the surface, which results in the subband structure of 2DES. Unlike 2DES formed in semiconductors, SE on liquid helium are free from any crystallographic defects and impurities of the substrate. Due to the pristine nature of this system,  interest in using quantum states of SE as quantum bits has been recently growing~\cite{PlatzDykm1999,Lea2000,DykmPRB2003,LyonAPL2008,SchuPRL2010,GePRX2016,KoolNatCom2019}. In particular, spin states of SE are expected to have longer coherence times than in any other solid-state materials~\cite{LyonPRA2006}. Although directly accessing the spin states of SE in an experiment is considered challenging~\cite{LyonPRA2006}, this task can be greatly facilitated by utilizing the coupling between the electron spin state and the electron motional state~\cite{SchuPRL2010,KawaPRL2019}. Therefore, elucidating the mechanism which governs the relaxation and dephasing of the  motional states of SE is of fundamental importance.

It has been theoretically demonstrated that the relaxation and dephasing of the electron motional states happen due to the elastic and inelastic scattering of SE from the excitation of liquid helium~\cite{Monarkha-book,DykmPRB2003,AndoJPSJ1978,SaitJPSJ1978,MonaLTP2006,MonaJLTP2007,MonaLTP2010}. There are two distinct liquid helium temperature ranges in this system. For high liquid helium temperatures $T$, the scattering is dominated by helium vapor atoms above the liquid ~\cite{AndoJPSJ1978,SaitJPSJ1978}. For $T$ sufficiently below 1~K, where the  concentration of vapor atoms becomes small, the scattering is mostly due to the capillary waves (ripplons) excited on the surface of the liquid~\cite{SaitJPSJ1978,MonaLTP2006,MonaJLTP2007,MonaLTP2010}. The rates of both the elastic scattering by vapor atoms and that by ripplons decrease with $T$ and become extremely small in the milli-Kelvin range. Such mechanisms of elastic scattering were experimentally established by measuring the electron mobility and coincide well with theory~\cite{AndoJPSJ1978}. Contrary to that, much less is known about the inelastic ripplon scattering rate, which governs the dissipative processes in SE at low $T$, such as the energy relaxation of the  motional states. In terms of using the Rydberg states as qubit states, studying the mechanism of the inelastic scattering process is important because it directly affects the relaxation and dephasing of the Rydberg states~\cite{AndoJPSJ1978,DykmPRB2003,MonaLTP2006,MonaJLTP2007,MonaLTP2010}. 

Some efforts on extracting the inelastic scattering rate from indirect measurements had been made, but it turned out to be  difficult.
The intrinsic spectral linewidth of the Rydberg states should be determined by the inelastic scattering rate at low $T$~\cite{CollPRL2002}. However, the inhomogeneous broadening, which is several orders of magnitude larger than the inelastic scattering rate, hinders us from measuring it with good precision ~\cite{CollPRL2002,IsshJLTP2007}.

The effective electron temperature $T_e$ should be also determined by the inelastic scattering rate via the energy balance equation~\cite{Monarkha2007-wf}. $T_e $ can be extracted from the mobility measurement under microwave (MW) radiation~\cite{SaitJPSJ1978,BadrEPL2013}. However, it relies on a complicated relationship between the electron temperature $T_e$ and the mobility of SE and thus is questionable.  An improved method to measure $T_e$, which is based on the thermoelectric (Seebeck) effect, was recently demonstrated~\cite{LyonPRL2018}. However, at present, this method is limited to the high-$T$ regime.  

In this manuscript, we report the first direct observation of the relaxation of the Rydberg state population of SE on liquid $^3$He in a wide range of $T$. This observation becomes possible thanks to the significant improvements in  image-charge detection~\cite{Asem-paper}. The relaxation times extracted from the decay signals show a cross-over between the vapor-atom and ripplon scattering regimes, and reach the maximum value of about 1~$\mu$s at the lowest $T=135$~mK used in the experiment. By comparing the measured signals and our theoretical calculations, we attribute this time to the inelastic scattering due to the spontaneous two-ripplon emission process. Moreover, we demonstrated that, at sufficiently low $T$, the image-charge signals become saturated shortly after applying the microwave radiation due to the saturation of the population between the two lowest states. The signals eventually increase due to the subsequent leaking of the SE population to the higher excited Rydberg states.      

SE are formed on the surface of the liquid $^3$He, which is set
approximately midway between two circular electrodes of a parallel-plate capacitor with capacitance $C_0$. As SE are excited from the ground ($n=1$) state to the higher ($n\geq 2$) Rydberg states, they cause a change in the image-charge in the electrodes~\cite{KawaPRL2019}. Whereas in the previous experiment the time-averaged image current was measured using a lock-in amplifier~\cite{KawaPRL2019}, in this work we measure a real-time image-charge response by employing a two-stage cryogenic broadband amplifier~\cite{Asem-paper}. The first stage, which is located close to the capacitor, converts the image current into a voltage signal of which the upper bound of the bandwidth exceeds 100~MHz. The total gain of our circuit is 1.  Thus both the voltage signal amplitude before amplification and the measured voltage signal can be written as $\Delta q/(C_0+C_p)$, where $C_p\sim 10$~pF is the total parasitic capacitance of our circuit and $\Delta q$ ($-\Delta q$) is the change in the image-charge in the top (bottom) capacitance plate due to excitation of SE to higher Rydberg states. As was shown previously~\cite{KawaPRL2019}, the image-charge change can be represented as      
 
\begin{equation}
\Delta q=\frac{e n_s C_0}{\epsilon_0} \sum\limits_n \left(\langle z \rangle_{nn} - \langle z \rangle_{11}\right)  \rho_{nn},
\label{eq:1}
\end{equation}

\noindent where $e$ is the elementary charge, $n_s$ is the density of SE, $\epsilon_0$ is the vacuum permittivity, $\langle z \rangle_{nn}$ is the average electron coordinate perpendicular to the surface of liquid for an electron occupying $n$-th Rydberg states, and $\rho_{nn}$ is the fractional occupancy of the $n$-th state. From the above equations, it is clear that the real-time voltage signal can capture the excited-state population dynamics in SE, providing that the detection bandwidth is large enough compared to the excitation and relaxation rates in the system.

\begin{figure}
\includegraphics[width=8.5cm]{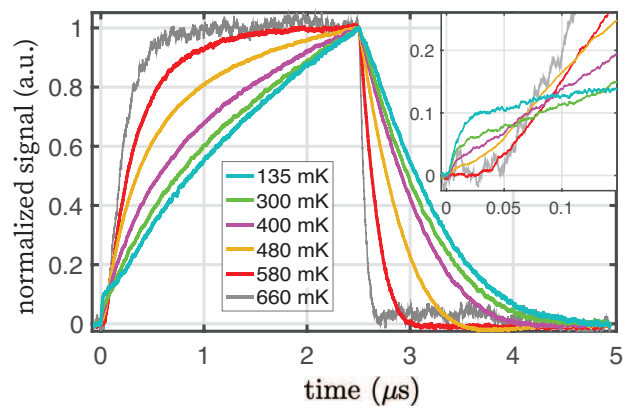}
\caption{(color online) Normalized voltage signals due to excitation of SE by pulsed MW radiation at frequency $\omega/2\pi=110$~GHz measured for different liquid helium temperatures $T$ in the range from 135~mK to 660~mK. The radiation is switched on for $0 \leq t \leq 2.5$~$\mu$s and is switched off for $2.5 \leq t \leq 5$~$\mu$s. Each trace is obtained at a fixed value of $T$ by averaging over about $10^6$ repetitions. The inset shows a magnified view of the main figure in a region of $t$ shortly after the radiation was switched on. The details of this region are discussed in the text.}
\label{fig:1}       % Give a unique label
\end{figure} 

In the experiment, SE are excited by the pulse-modulated (200~kHz) MW radiation at the frequency $\omega/2\pi=110$~GHz, which is resonant with the $n=1\rightarrow 2$ transition. The amplified voltage signals are averaged over approximately $10^6$ repetitions using a digital storage oscilloscope. The recorded traces are corrected to eliminate the effect of high-pass filtering in the detection circuit~\cite{SM}. Figure~\ref{fig:1} shows an example of normalized voltage signals obtained for different $T$ in the range from 135~mK to 660~mK. 

First, let us investigate how the overall signals differ depending on the liquid helium temperature $T$. During the MW excitation ($0<t<2.5 ~\mu s$), the voltage signals grow.  At sufficiently high $T$, they reach a constant value, which corresponds to the steady-state population of the Rydberg states. After MW radiation is turned off at $t=2.5~\mu$s, the voltage signal drops back to zero due to the relaxation of the excited-state population of SE  to the ground state. 

It is well established that SE can be strongly overheated by a resonant MW excitation, which causes a significant thermal population of the higher-lying Rydberg states~\cite{VolJETP1981,KonstPRL2007}. The heating is facilitated by fast elastic scattering of the MW-excited electrons between the subbands, such that the energy absorbed by MW excitation is  transferred into the energy of electron in-plane motion. Moreover, electrons redistribute energy within the same subband due to electron-electron collisions, such that SE are described by an effective electron temperature $T_e$, which can significantly exceed the liquid helium temperature $T$~\cite{KonstPRL2009,MonaJLTP2007,KonstPRB2012}. The electron-electron collision rate is of the same order as the plasmon frequency of SE and is much higher than any rate involved here \cite{KonstJPSC2008,Crandall1973-wm}. As a result, SE are thermally populated to excited states, although the only transition induced by the applied MW radiation is between the ground state and the first excited state. As the MW radiation is switched off, the decay of the excited-state population is governed by the electron energy dissipation due to inelastic scattering, upon which the energy of SE is transferred to the liquid helium excitation. 

In order to analyze the measured signals and extract the inelastic scattering rate, we performed  numerical simulations and calculated the image-charge signals, taking into account the elastic and inelastic scattering processes due to both the vapor-atoms and ripplons. We solve the coupled time-dependent energy-balance and rate equations to find the effective electron temperature $T_e(t)$ and fractional occupancy of the $n-$th Rydberg state $\rho_{nn}(t)$ for the pulse-modulated MW excitation~\cite{SM}. The rate of MW excitation is taken in the Lorentzian form $r=0.5\gamma\Omega_R^2/((\Delta\omega)^2+\gamma^2)$, where $\Delta\omega$ is the detuning from the resonance, $\gamma$ is the transition linewidth, and $\Omega_R$ is the Rabi frequency which is determined by the applied MW power~\cite{Loudon-book,AndoJPSJ1978,KonstJPSC2008}. For a given temperature $T$, the linewidth $\gamma$ is taken to be either the experimentally observed inhomogeneous linewidth or the calculated $T$-dependent intrinsic linewidth~\cite{AndoJPSJ1978}, whichever is largest. The intersubband transition rates, which enter into the rate equations for occupancies $\rho_{nn}$, are calculated by taking into account the elastic scattering due to both vapor-atoms and ripplons~\cite{KonsFNT2008}. The electron energy loss rate, which enters into the energy-balance equation, are calculated by taking into account the inelastic scattering from both vapor-atoms and ripplons~\cite{KonsFNT2008}. Following ~\cite{MonaFNT1978,MonaLTP2006,MonaJLTP2007,MonaLTP2010}, we calculate the inelastic scattering due to ripplons by assuming a finite value of the surface potential barrier and taking the second-order term in the ripplon-induced surface displacement in the interaction Hamiltonian. By doing so, we approximate the inelastic collisions between electrons and ripplons to the two-ripplon scattering processes. Finally, the time-dependent image-charge signal for a given $T$ is calculated as $\sum_n \left(\langle z \rangle_{nn} - \langle z \rangle_{11}\right)  \rho_{nn}(t)$.                              

Figure 2 shows the normalized image-charge signals calculated for several temperatures in the range from 135~mK to 580~mK and a fixed value of the Rabi frequency $\Omega_R/2\pi=40$~MHz. As in the experiment, the MW excitation is present at $0\leq t \leq 2.5$~$\mu$s and is absent at $2.5\leq t \leq 5$~$\mu$s. From the comparison with the normalized measured signals shown in Fig.~\ref{fig:1}, we conclude that the simulations account quite well for the observed behavior. It is important to emphasize that our simulations do not contain adjustable parameters except the chosen value of the Rabi frequency $\Omega_R$, which  is determined by the applied MW power in  the experiment. 

Second, let us concentrate attention on the decay signal observed after the MW radiation is turned off. Figure~\ref{fig:4}(a) shows the semi-log plot of the beginning of the decaying part of the normalized measured signals shown in Fig.~\ref{fig:1}. Although the decay does not follow a simple exponential law, in order to quantify the decay rate we fit a part of the signal with the analytical expression $\exp\left( -t/\tau_f\right)$, where $\tau_f$ is a fitting parameter. The fitting curves are shown with dashed lines for each signal in Fig.~\ref{fig:4}(a). The values of $\tau_f$ extracted from the decaying signals for each $T$ are plotted in Fig.~\ref{fig:4}(b) as a function of $T$ in the semi-log plot by filled (black) circles. As $T$ decreases, the decay time $\tau_f$ rapidly increases for $400\lesssim T \leq 700$~mK. In this temperature range, the decay rate is governed by the inelastic scattering of SE from the vapor-atoms, whose concentration decreases with decreasing $T$ according to an exponential law $\exp (-Q/T)$, where $Q$ is the latent heat of evaporation ($Q\approx 2.47$~K for $^3$He~\cite{SaitJPSJ1978}). For $135\leq T\lesssim 400$~mK, the decay time $\tau_f$ tends to saturate with decreasing $T$. The longest measured decay time is $\tau_f\approx 0.8$~$\mu$s at $T=135$~mK. At low $T$, the concentration of vapor atoms becomes so small that the inelastic scattering is dominated by the ripplons. According to the theory~\cite{MonaFNT1978,MonaLTP2006,MonaJLTP2007,MonaLTP2010}, the main mechanism of the electron energy loss due to the ripplons is the spontaneous two-ripplon emission process. The dependence of this process on $T$ is much weaker than the inelastic scattering from vapor-atoms.

\begin{figure}
\includegraphics[width=8.5cm]{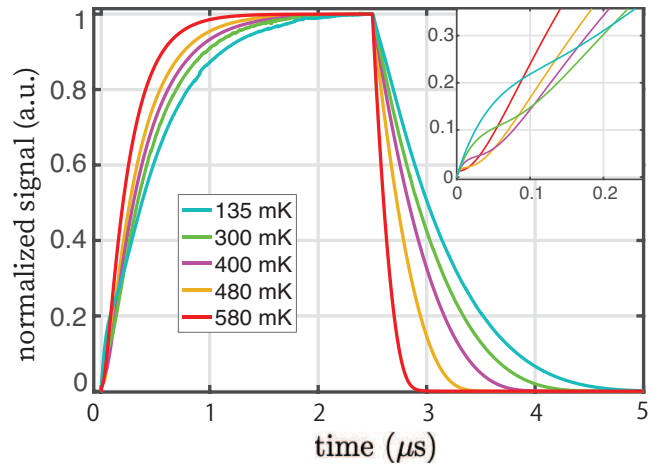}
\caption{(color online) Normalized image-charge signals (in arbitrary units) due to excitation of SE by pulsed microwave radiation calculated using the excited-state occupancies $\rho_{nn}$ obtained by numerically solving the rate and energy balance equations~\cite{SM}. In the calculations, the radiation power corresponds to the Rabi frequency $\Omega_\mathrm{R}/2\pi=40$~MHz. For the sake of comparison with Fig.~\ref{fig:1}, the inset shows a magnified view of the main figure in a region of $t$ shortly after the excitation is started.}
\label{fig:2}       % Give a unique label
\end{figure}

According to Ref.~\cite{KonsFNT2008} the electron energy loss rate $\dot{E}$ can be represented as $-(\tilde{\nu}_{r}+\tilde{\nu}_a) (T_e-T)$, where $\tilde{\nu}_{r}$ and $\tilde{\nu}_a$ are the effective energy relaxation rate for ripplon and vapor-atom scattering, respectively, which depends in a rather complicated way on the electron temperature and state occupancies. Nevertheless, it is still instructive to compare the signal-decay rates obtained from the above fitting procedure with the rates $\tilde{\nu}_{r}$ and $\tilde{\nu}_a$ evaluated at a fixed value of $T_e$. Such numerical evaluation can be easily done by assuming  thermal state occupancies according to the Boltzmanm distribution~\cite{SM}. In Fig.~\ref{fig:4}(b), the inverse rates $\tilde{\nu}_{r}^{-1}$, $\tilde{\nu}_a^{-1}$ and $(\tilde{\nu}_{r}+\tilde{\nu}_a)^{-1}$ evaluated at the stationary value of $T_e$ for SE excited by resonant MW radiation with $\Omega_R/2\pi=40$~MHz are plotted by the dotted, dashed and solid lines, respectively. The temperature $T\approx 550$~mK corresponds to the cross-over between the inelastic scattering regime dominated by ripplons and the one dominated by vapor-atoms. 

For the sake of comparison with the simulations, in Fig.~\ref{fig:4}(b) we also plot the decay time $\tau_f$. These times were extracted by the same fitting procedure from the simulated decaying signals and were calculated for three different values of the Rabi frequency $\Omega_R/2\pi=30$ (blue pointing-down triangles), 40 (purple rectangles) and 50~MHz (green pointing-up triangles). We found that the values of $\Omega_R$ in this range provide the best correspondence between the experimentally observed and theoretically calculated decay rates. We note that, although the exact relationship between the applied MW power and the Rabi frequency can not be accurately determined in the experiment, the above values of the Rabi frequency are in  good agreement with the earlier estimations of $\Omega_R$ obtained from the non-linear conductivity of SE on liquid $^3$He under the resonant MW excitation~\cite{KonstPRL2007}.

\begin{figure}[h!]
	\centering
	\includegraphics[width=1.0\linewidth]{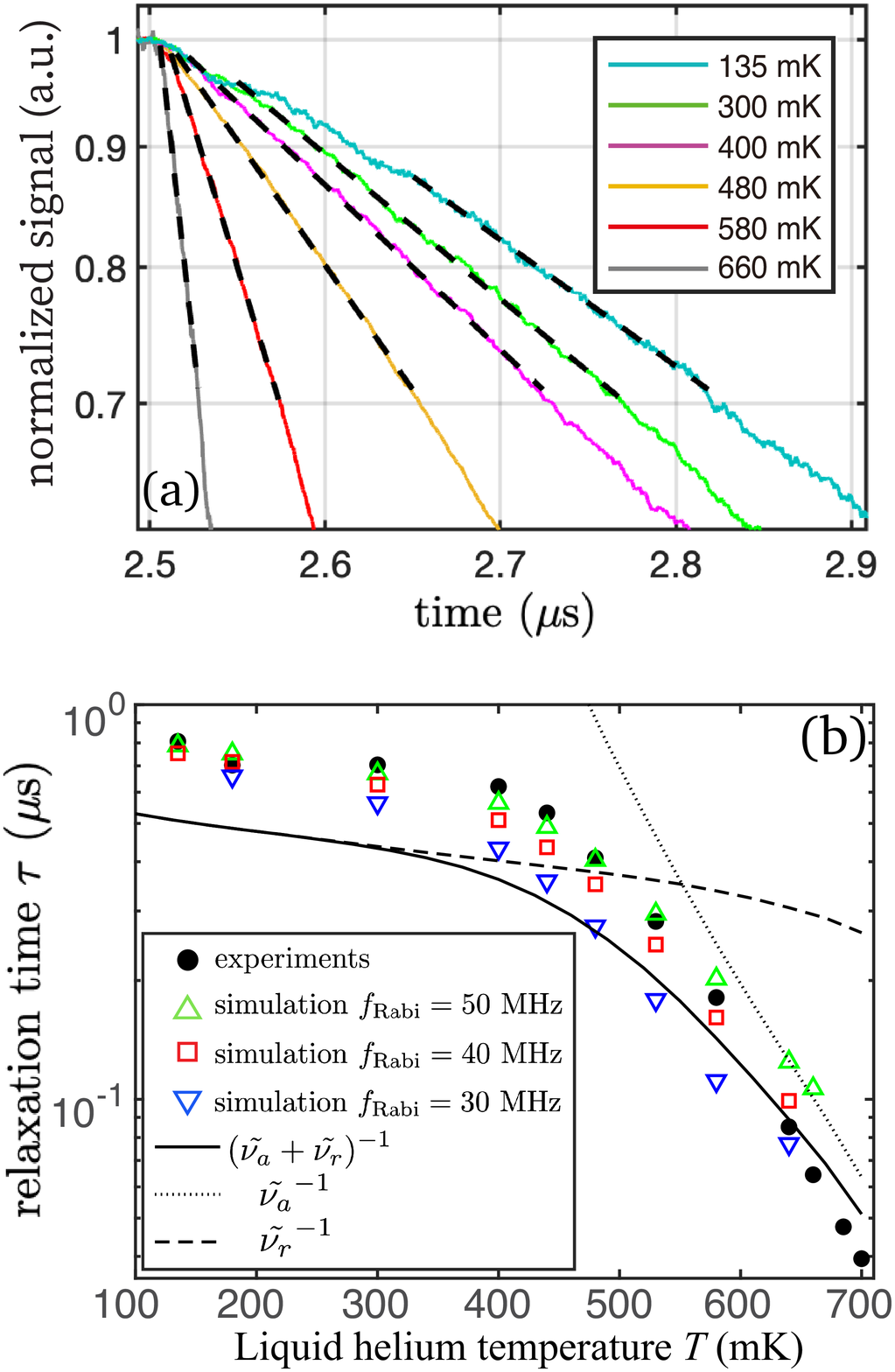}
	\caption{(color online) (a) Semi-log plot of the decaying voltage signals shown in Fig.~\ref{fig:1} after MW is switched off. Dashed lines are fits with the exponential decay law $\exp(-t/\tau_f)$, where $\tau_f$ is the fitting parameter. (b) The fitting parameter $\tau_f$ (filled black circles) plotted as a function of liquid helium temperature $T$. For the sake of comparison, the relaxation times $\tau_f$ extracted from the similar fitting of the numerically calculated image-charge decay curves for different values of the Rabi frequency $\Omega_\mathrm{R}/2\pi=30$ (blue pointing down triangles), 40 (purple unfilled rectangles), and 50~MHz (green pointing up triangles) are also plotted. The lines represent the inverse of the energy relaxation rates $\tilde{\nu}$ calculated according to Ref.~\cite{KonsFNT2008} for SE with the electron temperature $T_e$, which was obtained by solving the stationary rate and energy balance equations for $\Omega_\mathrm{R}/2\pi=40$~MHz~\cite{SM}, by taking into account the inelastic vapor-atom scattering (dashed line), the inelastic two-ripplon scattering (dotted line), and both processes simultaneously (solid line).}
	\label{fig:4}
\end{figure}

Finally, we address an interesting behavior of the measured signals shortly after the MW excitation is started, see the inset of Fig.~\ref{fig:1}. Remarkably, this rather complicated behavior is reproduced well by our numerical simulations, c.f. the inset of Fig.~\ref{fig:2}. By analyzing the calculated state occupancies~\cite{SM}, we conclude that this behavior corresponds to the initial saturation of the first-excited state population due to MW excitation and subsequent population of the higher-lying Rydberg states. At sufficiently low $T\lesssim 400$~mK, when the MW-excitation rate of SE exceeds the rate of elastic scattering, the saturation of the first-excited state causes a plateau in the measured signal. As time progresses, the elastic scattering facilitates the heating and thermal population of the higher-excited states; therefore, the signal grows towards a value corresponding to the steady-state population of many subbands. At high $T$, when the rate of elastic scattering exceeds the excitation rate, electrons start populating the higher-lying Rydberg states soon after applying the excitation; therefore the signal steadily grows towards the stationary value. We note that this $T$-dependent behavior can be reproduced only for a certain range of the values of $\Omega_R$, which define the MW-excitation rate of SE. Thus, the good agreement between the experimental data and calculations provide strong support for our estimation of the Rabi frequency in the experiment.               

In summary, we described the first direct observation of the relaxation of the excited-state population in a multi-subband electron system on liquid helium. This allowed us to establish the temperature-dependent inelastic scattering mechanisms in the system, which will determine the quality of qubits when the Rydberg states of electrons are used as quantum bits~\cite{DykmPRB2003}. In particular, our experimental results are well captured by the energy dissipation mechanism due to spontaneous two-ripplon emission~\cite{MonaLTP2006,MonaJLTP2007,MonaLTP2010,MonaFNT1978}. We also demonstrated that, while the elastic scattering makes higher-lying Rydberg states populated, the population of the two lowest states,  the transition between which is induced by MW radiation, becomes saturated only a short time after MW radiation is started. We note that, as was proposed earlier~\cite{DahmJLTP2002}, the population of higher-lying Rydberg states can be completely suppressed by applying a sufficiently strong ($\sim 5$~T) magnetic field perpendicular to the surface of the liquid. Thus, our method of fast image-charge detection can provide a valuable tool to study the two-level population dynamics in SE at high magnetic fields.

This work was supported by JST-PRESTO (Grant No. JPMJPR1762) and an internal grant from Okinawa Institute of Science and Technology (OIST) Graduate University.

%%%%%%%%%% Merge with supplemental materials %%%%%%%%%%
\pagebreak
\widetext
\begin{center}
\textbf{\large Supplemental Materials: Relaxation of the Excited Rydberg States of Surface Electrons on Liquid Helium}
\end{center}
%%%%%%%%%% Merge with supplemental materials %%%%%%%%%%
%%%%%%%%%% Prefix a "S" to all equations, figures, tables and reset the counter %%%%%%%%%%
\setcounter{equation}{0}
\setcounter{figure}{0}
\setcounter{table}{0}
\setcounter{page}{1}
\makeatletter
\renewcommand{\theequation}{S\arabic{equation}}
\renewcommand{\thefigure}{S\arabic{figure}}
\renewcommand{\bibnumfmt}[1]{[S#1]}
\renewcommand{\citenumfont}[1]{S#1}
%%%%%%%%%% Prefix a "S" to all equations, figures, tables and reset the counter %%%%%%%%%%

\section{Effect of high-pass filtering in the detection circuit}

\begin{figure}[b]
\includegraphics[width=15cm]{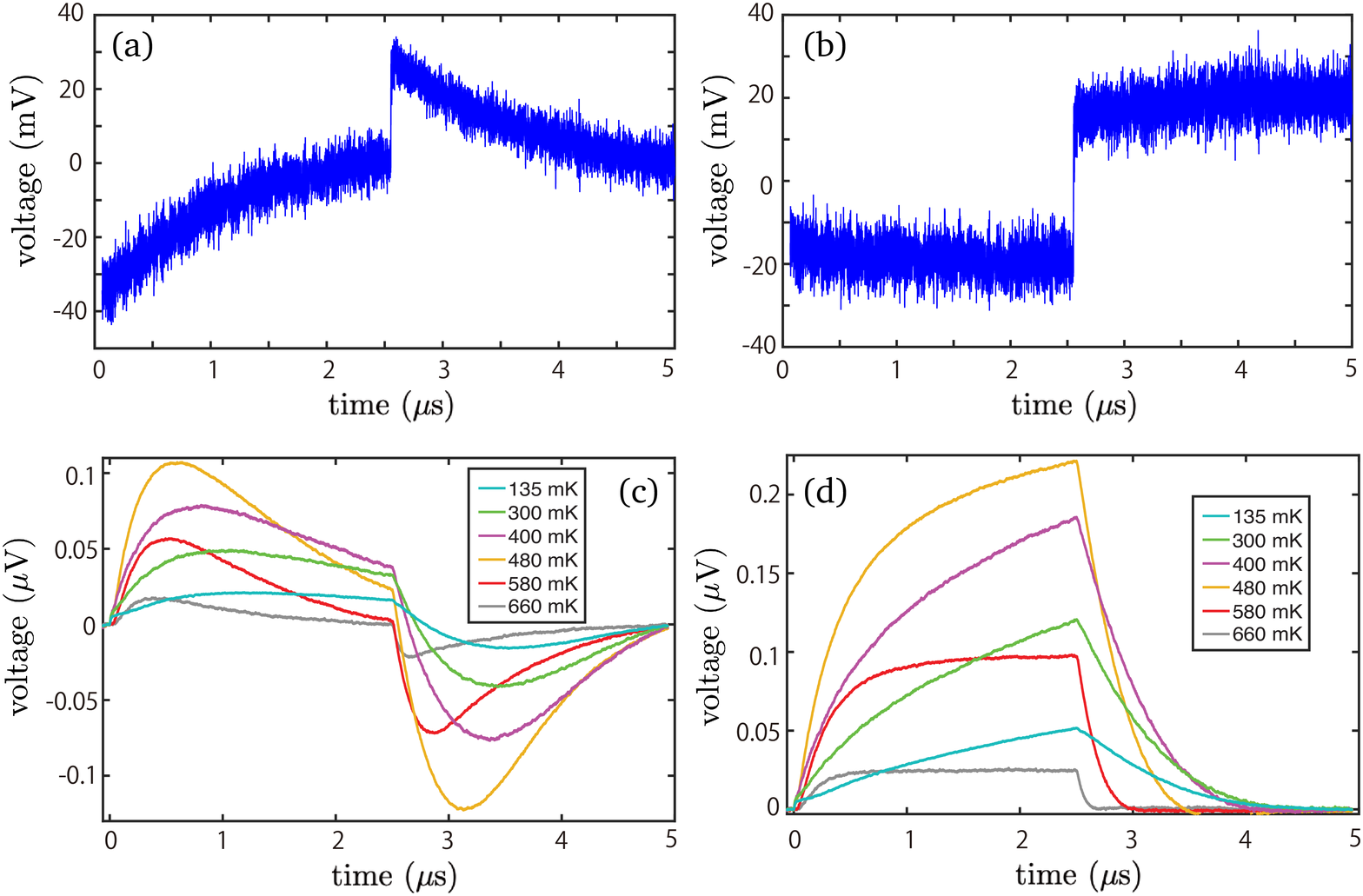}
\caption{(color online)  (a) Voltage signal measured at the output of the two-stage amplifier without electrons and when a square wave with  amplitude 20~mV and frequency 200~kHz is applied to the capacitor bottom plate. (b) Corresponding voltage signal at the capacitor top plate obtained by applying the recurrence relation of the high-pass filter to the signal shown in panel (a). (c) Voltage signals measured at the output of the two-stage amplifier due to the surface electrons excited by pulse modulated MW radiation. (d) Corresponding voltage signals at the capacitor top plate obtained by applying the recurrence relation of the high-pass filter the voltage signals shown in panel (c).}
\label{fig:S1}       % Give a unique label
\end{figure} 

As described in the main text, the signal induced by the excited surface electrons at the top plate of the capacitor was detected using a two-stage cryogenic broadband amplifier~\cite{Asem-paper_S}. Prior to the measurements of the signal, the time domain response of the detection circuit was checked by applying a square wave with  amplitude 20~mV and frequency 200~kHz to the bottom plate of the capacitor. Fig.~\ref{fig:S1}(a) shows the corresponding signal measured at the output of the amplifier at $T\sim 100$~mK. From the analysis of this time domain response we conclude that the frequency response of the two-stage amplifier circuit can be approximated as a second-order high pass filter where the cut-off frequency of the first stage is 135~kHz and that of the second stage is 53~kHz. By applying the recurrence relation of the high-pass filter in an inverse way to the measured signal (Fig.~\ref{fig:S1}(a)), we recover the time domain signal induced at the capacitor bottom plate (Fig.~\ref{fig:S1}(b)). Using this procedure, we recover the voltage signals at the capacitor top plate due to the surface electrons. Fig.~\ref{fig:S1}(c) shows the signals measured at the output of the amplifier due to electrons excited by the pulsed microwave (MW) radiation at different temperatures $T$, as described in the main text. Fig.~\ref{fig:S1}(d) shows the corresponding voltage signals induced at the bottom capacitor plate, which were recovered using the above procedure. The corresponding traces, normalized for each value of $T$, are shown in Fig.~1 of the main text.

\section{Numerical simulation of the image-charge signals due to excitation of surface electrons}

In order to obtain the image-charge signals due to pulse-modulated MW excitation of surface electrons, we numerically solved the coupled time-dependent energy-balance and rate equations to find the electron temperature $T_e$ and the Rydberg state occupancies $\rho_{nn}$ by using the Ordinary Differential Equation (ODE) solver in Matlab. The energy-balance equation can be written as (hereafter, for the sake of simplicity we assume the Boltzmann constant $k_\textrm{B}=1$)

\begin{equation}
\frac{dT_e}{dt}+\sum_n \Delta_{n1} \frac{d\rho_{nn}}{dt}= \hbar \omega r (\rho_{11}-\rho_{22}) + \dot{E}, 
\label{eq:ebalance}
\end{equation}

\noindent where we use a notation $\Delta_{n'n}=E_{n'}-E_n$ for the energy difference between the Rydberg states with indexes $n'$ and $n$. The MW excitation rate for an electron at the resonance is given by $r=0.5\Omega^2/\gamma$, as described in the main text. The left-hand side of Eq.~(\ref{eq:ebalance}) represents the time derivative of the mean total energy of an electron, which is sum of the kinetic energy of electron in-plane motion and the quantized energy of the out-of-plane motion. The first term in the right-hand side of Eq.~(\ref{eq:ebalance}) represents the rate of energy absorption by an electron from the MW field. The second term is the rate of energy loss by an electron due to inelastic scattering, which can be written as the sum of the energy loss rate due to the vapor-atom scattering \cite{SaitJPSJ1978_S} and the energy loss rate due to the two-ripplon inelastic scattering~\cite{KonsFNT2008_S,MonaLTP2010_S} 

\begin{align}
\dot{E}=& -\frac{\pi \hbar A N_G}{M} \left( 1-\frac{T}{T_e}\right) \sum_{nn'} \rho_{nn} e^{-\frac{\left| \Delta_{n'n}  \right|+\Delta_{n'n}}{2T_e}} \left[(2T_e+\left| \Delta_{n'n}  \right|)  \int_0^\infty (\psi_{n'} (z)\psi_n (z) )^2 dz + \frac{\hbar^2}{2m}  \int_0^\infty \frac{d}{dz} (\psi_{n'} (z)\psi_n (z) )^2 dz \right]
\nonumber \\
& -\frac{m}{2\pi \rho^2} \sum_{nn'} \rho_{nn} \left|\frac{1}{2}  \Bigg \langle n \left| \frac{\partial^2 V_e^{(0)}}{\partial z^2}  \right| n' \Bigg \rangle  \right|^2 \int_0^\infty dq \frac{q^3}{\omega_q} (N_q+1)^2 \left( 1-\frac{\rho_{n'n'}}{\rho_{nn}} e^\frac{\Delta_{n'n}}{T_e} e^{2\hbar\omega_q\left(\frac{1}{T_e} -\frac{1}{T} \right)} \right) e^{-\frac{\left| \Delta_{n'n} + 2\hbar\omega_q \right| + \Delta_{n'n} + 2\hbar\omega_q }{2T_e}}.
\label{eq:erate}
\end{align}

\noindent Here, $N_G$ is the concentration of the vapor atoms, $A=4.98\times 10^{-16}$~cm$^2$ is the cross-section of a helium atom~\cite{AndoJPSJ1978_S}, $m$ and $M$ is the mass of an electron and a helium atom, respectively, $\rho$ and $\alpha$ is the density and the surface tension of liquid helium, respectively, $\psi_n (z)$ is the electron wave-function corresponding to the $n$-th Rydberg state, $N_q$ is the mean occupation number of ripplons having the usual capillary wave dispersion relation $\omega_q=\sqrt{\alpha q^3/\rho}$ with wave number $q$ of ripplons, and $V_e^{(0)}$ is the potential energy of an electron over the flat surface of liquid helium. Note that the energy loss rate due to the one-ripplon inelastic scattering is much smaller than that given by Eq.~\eqref{eq:erate} in the whole temperature range, therefore is omitted.

The balance equations for the rates of the inter-subband transitions of an electron can be written as

\begin{align}
    \frac{d \rho_{nn}}{dt}=&\sum_{n' \neq n} (\nu_{n'n}\rho_{n'n'}- \nu_{nn'}\rho_{nn} ),  \mathrm{         \ \ \ \  }  (n \geq 3)
 \nonumber \\
\frac{d \rho_{22}}{dt}=& r (\rho_{11} - \rho_{22}) +\sum_{n' \neq 2} (\nu_{n'2}\rho_{n'n'}- \nu_{2n'}\rho_{22} ), \nonumber \\
\frac{d \rho_{11}}{dt}=& r (\rho_{22} - \rho_{11}) +\sum_{n' \neq 1} (\nu_{n'1}\rho_{n'n'}- \nu_{1n'}\rho_{11} ), 
\label{eq:rbalance}
\end{align}

\noindent In the above equations, the inter-subband transition rates $\nu_{n'n}$ are dominated by the elastic scattering of an electron from the gas vapor-atoms and ripplons, and can be written as~\cite{KonsFNT2008_S,MonaLTP2010_S}

\begin{equation}
\nu_{n'n}= \frac{\pi \hbar N_G A}{m} e^{-\frac{\left| \Delta_{n'n}  \right| +\Delta_{n'n}}{2 T_e}} \int_0^\infty  \left[   \psi_{n'}(z) \psi_{n}(z) \right]^2 dz   + \frac{T}{4 \sqrt{\pi}  \alpha \hbar \sqrt{T_e} }  \int_0^\infty \frac{ d\epsilon_q }{\epsilon_q^{3/2} } \left| \langle n \left| U_q(z)  \right| n'  \rangle  \right|^2 e^{-\frac{\left( \epsilon_q + \Delta_{n'n}  \right)^2  }{4 \epsilon_q T_e}},
\label{eq:nu}   
\end{equation}

\noindent where $U_q(z)$ is the electron-ripplon coupling~\cite{KonsFNT2008_S,MonaLTP2010_S} and, for the sake of convenience, we introduced notation $\epsilon_q=\hbar^2q^2/(2m)$.  The inter-subband transition rates due to the inelastic two-ripplon scattering become non-negligible only when $T \lesssim 10$~mK, which is outside the temperature range considered here.

In order to obtain the results shown in Fig.~2 of the main text, the 10 lowest Rydberg states where included in the simulations. To decrease the computation time, only the 5 lowest Rydberg states were included to obtain the data shown in Fig.~3(b) of the main text. We have confirmed the convergence of the results for both cases.

\section{Stationary electron temperature of the microwave-excited electrons}

\begin{figure}
\includegraphics[width=7cm]{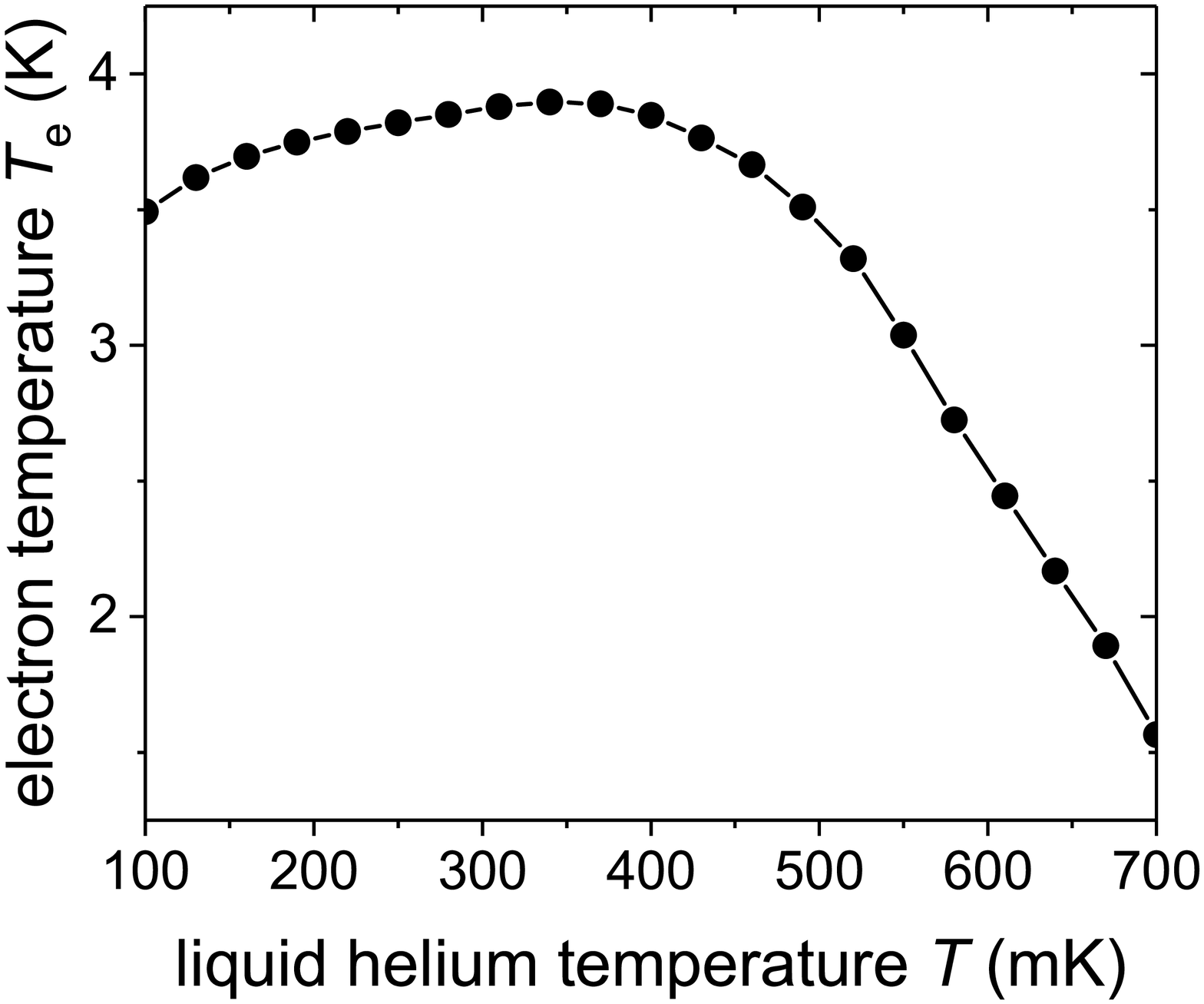}
\caption{(color online)  Stationary value of $T_e$ for the electrons excited by the resonant microwave radiation with intensity corresponding to the Rabi frequency $\Omega=40$~MHz for different values of the liquid helium temperature $T$.}
\label{fig:S2}       % Give a unique label
\end{figure}

The stationary value of $T_e$ can be found from Eqs.~(\ref{eq:ebalance},\ref{eq:rbalance}) by equating the time derivatives of $T_e$ and $\rho_{nn}$ to zero. The values of $T_e$ calculated for the Rabi frequency $\Omega=40$~MHz at different values of $T$ are shown in Fig.~\ref{fig:S2}. The 10 lowest Rydberg states were included in these calculations. 

For the sake of comparison with the experimentally observed rate of signal decay, we can estimate the rate of electron energy relaxation due to scattering by the helium vapor-atoms and ripplons, $\tilde{\nu}_a$ and $\tilde{\nu}_r$, respectively, by representing the electron energy loss rate as $\dot{E}=-(\tilde{\nu}_a+\tilde{\nu}_r)(T_e-T)$. For an estimation of $\tilde{\nu}_a + \tilde{\nu}_r$, we calculate the energy loss rate $\dot{E}$ using Eq.~\eqref{eq:erate} and assuming the thermal (Boltzmann) distribution of electrons over the Rydberg states corresponding to the stationary value of $T_e$. The results are plotted with lines in Fig.~3(b) of the main text.

\section{Subband occupancies for the microwave-excited electrons}

\begin{figure}[ht]
\includegraphics[width=15cm]{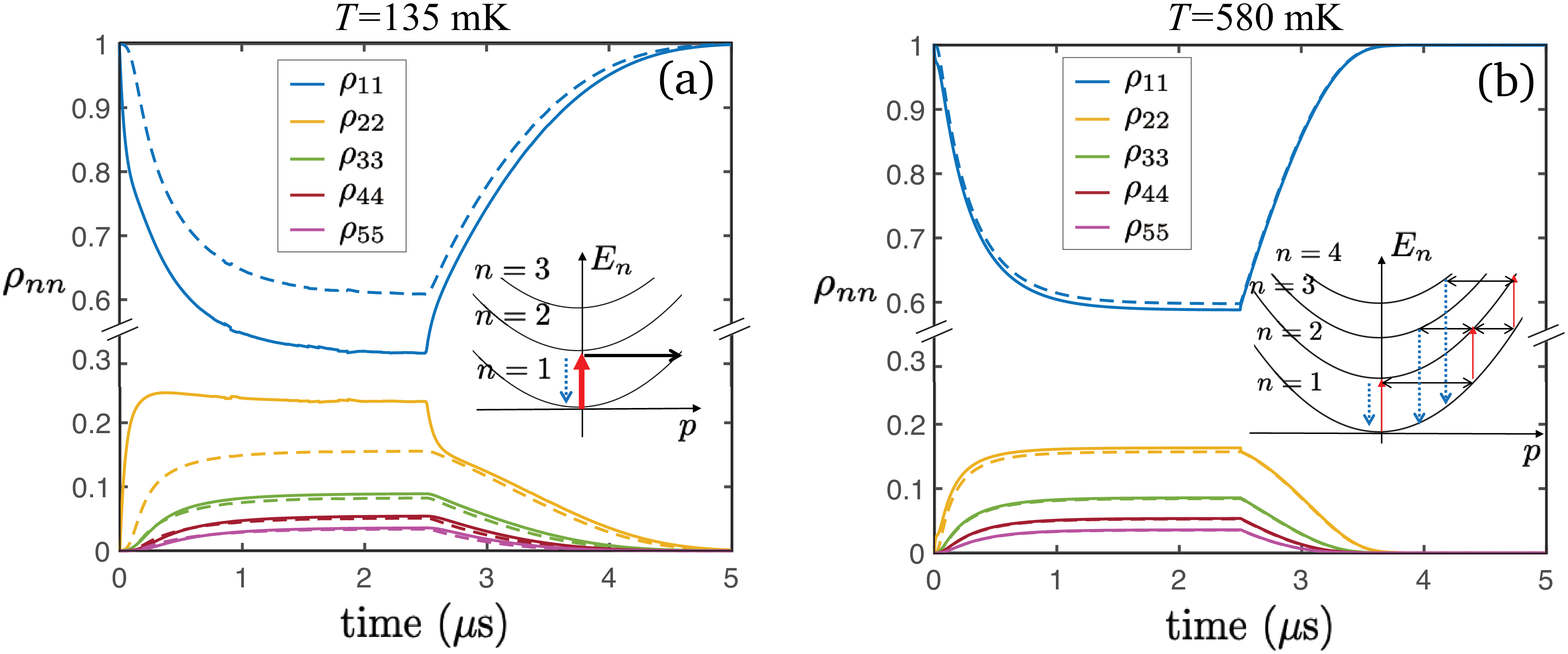}
\caption{(color online)  Fractional occupancies $\rho_{nn}$ for the 5 lowest Rydberg states calculated for pulse-modulated resonant MW excitation (excitation is turned on for $0\leq t \leq 2.5~\mu$s and excitation is turned off for $2.5\leq t \leq 5~\mu$s) for $\Omega=40$~GHz and for two values of $T=135$ (a) and 580~mK (b). The solid lines represent the occupancies obtained by numerically solving the coupled energy-balance and rate equations (\ref{eq:ebalance},\ref{eq:rbalance}). The dashed lines represent the thermal (Boltzmann) occupancies calculated using the value of $T_e$ obtained from Eqs.~(\ref{eq:ebalance},\ref{eq:rbalance}). Insets for each figure show sketches of the energy subbands and the inter-subband transitions of electrons shortly after MW excitation is applied. At $T=135~$mK, see the inset of (a), the MW-induced transitions (red arrow) are more dominant than the transitions induced by the elastic (black arrow) and inelastic (blue dotted arrow) scatterings. Correspondingly, the electron system exhibits saturation of the two lowest subbands. At $T=580~$mK, see the inset of (b), the transitions due to the elastic scattering (black arrows) is more dominant than the transitions due to the inelastic scattering (blue dotted arrows) and the MW-induced transitions (red arrows). Correspondingly, the electrons populate many subbands and the subband occupancies are close to the thermally determined ones.}
\label{fig:S3}       
\end{figure}

It is important to analyze the Rydberg state occupancies for the MW-excited electrons for different ranges of $T$. Fig.~\ref{fig:S3} shows exemplary plots of the state occupancies $\rho_{nn}$ for the 5 lowest Rydberg states calculated for pulse-modulated resonant MW excitation with $\Omega=40$~GHz and for two values of $T=135$ and 580~mK. The solid lines represent the occupancies obtained by numerically solving the coupled energy-balance and rate equations Eqs.~(\ref{eq:ebalance},\ref{eq:rbalance}), as described earlier. For the sake of comparison, the dashed lines represent the thermal (Boltzmann) occupancies calculated using the value of $T_e$ obtained from Eqs.~(\ref{eq:ebalance},\ref{eq:rbalance}). 

We find that the occupancies of the Rydberg states with $n\geq 3$ calculated by the two methods are very close to each other in the whole range of $T$ considered here. Contrarily, the occupancies $\rho_{11}$ and $\rho_{22}$ obtained by solving the rate equations deviate from the corresponding thermal occupancies at sufficiently low $T$, see Fig.~\ref{fig:S3}(a). This is due to the strong temperature dependence of the inter-subband transition rates $\nu_{nn'}$ which appear in Eq.~(\ref{eq:rbalance}). According to Eq.~(\ref{eq:nu}), these rates satisfy the condition $\nu_{n'n}=\nu_{nn’}\exp(-\Delta_{nn'}/T_e)$. When the MW excitation rate $r$ is sufficiently small, it can be neglected in Eq.~(\ref{eq:rbalance}) and $\rho_{nn}$ is expected to be close to the thermal occupancies for all $n$. For Rabi frequency $\Omega=40$~GHz, which determines the MW excitation rate $r$, this is the case for $T\gtrsim 300$~mK. At lower values of $T$, the inter-subband transition rates decrease, so the MW excitation rate cannot be ignored. As a result, the occupancies of the two lowest Rydberg states can significantly deviate from the thermal ones.

The deviation of the occupancies $\rho_{11}$ and $\rho_{22}$ from the thermally determined ones for the MW-excited electrons strongly affect the shape of the image-charge signals. First, we consider the image-charge signals shortly after MW is turned on, see the insets of Fig.~1 and Fig.~2 of the main text. At sufficiently low $T\lesssim 300$~mK, the signals increase rapidly and quickly saturate. From the analysis of the state occupancies we conclude that this is due to the fast population of the first excited ($n=2$) Rydberg state with the rate $r$ determined by the Rabi frequency $\Omega=40$~GHz. Since the MW-induced transition is a dominant process here, the system behaves like a two-level system for sufficiently short times after MW excitation is started, see inset of Fig.~\ref{fig:S3}(a). Contrarily, at sufficiently large $T\gtrsim 300$~mK, the dominant process is the elastic inter-subband scattering of electrons, which quickly distributes the MW-excited electrons over many subbands, see inset of Fig.~\ref{fig:S3}(b). Here, the state occupancies closely follow the Boltzmann distribution with the electron temperature $T_e$ and the image-charge signal  grows according to increasing $T_e$ from the very beginning of the MW excitation.

Finally, let us focus on the decaying part of the image-charge signals at $t\geq 2.5~\mu$s, see Figs.~1 and 2 of the main text. For low $T\lesssim 300$~mK, the steady-state occupancies $\rho_{11}$ and $\rho_{22}$ at $t=2.5~\mu$s strongly deviate from the thermal ones, see Fig.~\ref{fig:S3}(a). After MW excitation is turned off, the state occupancies $\rho_{11}$ and $\rho_{22}$ quickly change and approach the thermal ones due to the elastic inter-subband scattering. After all the subbands are set to follow the Boltzmann distribution, the decay rate is governed by the inelastic scattering. The quick change in $\rho_{11}$ and $\rho_{22}$ during the thermalization process is observed as a faster drop of the measured image-charge signals just after turning off MW excitation for $T$=135, 300, and 400~mK, see Fig.~3(a) of the main text. Therefore, the starting points of the fitting regions for each $T$ in Fig.~3(a) of the main text are selected such that electrons had sufficient time to thermalize, so that the decay of the image-charge signals is governed by the inelastic scattering.

\end{document}